\documentclass[aps,prb,reprint,showpacs,superscriptaddress]{revtex4-1}

\usepackage{amsmath,amssymb,revsymb}
\usepackage{graphicx}% Include figure files
\usepackage{dcolumn}% Align table columns on decimal point
\usepackage{bm}% bold math
\newcommand{\calH}{\mathcal{H}}

\newcommand{\calO}{\mathcal{O}}
\newcommand{\calP}{\mathcal{P}}

\newcommand{\vecd}{\bm{d}}

\newcommand{\veck}{\bm{k}}

\newcommand{\vecs}{\bm{s}}

\newcommand{\tr}{{\rm tr}\,}

\begin{document}

%\preprint{}
\title{
Non-magnetic impurity effects in a three-dimensional topological superconductor: From $p$- to $s$-wave behaviors}
%From $p$- to $s$-wave behaviors in a three-dimensional topological superconductor with non-magnetic impurities}

\author{Yuki Nagai}%
\affiliation{%
CCSE, Japan Atomic Energy Agency, 5-1-5 Kashiwanoha, Kashiwa, Chiba
 277-8587, Japan}
\author{Yukihiro Ota}%
\affiliation{%
CCSE, Japan Atomic Energy Agency, 5-1-5 Kashiwanoha, Kashiwa, Chiba
 277-8587, Japan}
\author{Masahiko Machida}%
\affiliation{%
CCSE, Japan Atomic Energy Agency, 5-1-5 Kashiwanoha, Kashiwa, Chiba
 277-8587, Japan}%
%\affiliation{%
%CCSE, Japan Atomic Energy Agency, 5-1-5 Kashiwanoha, Kashiwa, Chiba
% 277-8587, Japan}%
%\affiliation{%
%Computational Materials Science Research Team, RIKEN AICS, Kobe, Hyogo
%650-0047, Japan 
%}%

\date{\today}% It is always \today, today,
             %  but any date may be explicitly specified

\begin{abstract}
Unconventional features in superconductivity are revealed by responses to impurity scattering. 
We study non-magnetic impurity effects in a
 three-dimensional topological superconductor, focusing on an effective
 model (massive Dirac Hamiltonian with $s$-wave on-site pairing) of
 Copper-doped bismuth-selenium compounds.  
Using a self-consistent $T$-matrix approach for impurity scattering, we
 examine in-gap states in density of states. 
We find that the results are well characterized by a single material
 variable, which measures relativistic effects in the
 Dirac Hamiltonian. 
In non-relativistic regime, an odd-parity
 superconducting state is fragile against non-magnetic impurities. 
We show that this behavior is caused by a $p$-wave character involved in
 the topological superconducting state. 
In contrast, we show that in relativistic regime the superconductivity
 is robust against non-magnetic impurities, owing to an $s$-wave
 character. 
To summarize, the system has two aspects, $p$- and
 $s$-wave features, depending on the weight of relativistic effects. 
\end{abstract}

\pacs{74.20.Rp,74.25.-q,74.90.+n}% PACS, the Physics and Astronomy
                             % Classification Scheme.
%\keywords{Suggested keywords}%Use showkeys class option if keyword
                              %display desired
\maketitle

%\tableofcontents

\section{Introduction}
Topological materials~\cite{Prange;Girvin:1990,Nakahara:2003,Volovik:2003,Hasan;Kane:2010,Ando:2013} are 
attracting a great deal of attention in condensed matter physics.
The seminal
studies~\cite{Schnyder;Ludwig:2008,Fu;Berg:2009,Sato:2010,Sato;Fujimoto:2010} 
indicate that a superconducting state can emerge, with topological
invariants, such as the Thouless-Kohmoto-Nightingale-Nijs
invariant~\cite{Thouless;Nijs:1982} and the Kane-Mele $Z_{2}$
invariant.~\cite{Kane;Mele:2005,Fu;Kane:2007} 
Topological superconductivity is predicted in both two- and
three-dimensional systems. 
Copper-doped bismuth-selenium
compounds~\cite{Hor;Cava:2010,Wray;Hasan:2010} are candidates for 
bulk topological superconductors, and their properties
are studied by different physical probes, including 
point-contact spectroscopy,~\cite{Sasaki;Ando:2011,Kirzhner} magnetization
curve,~\cite{Das;Kadowaki:2011} and scanning tunneling
spectroscopy.~\cite{Levy;Stroscio:2013} 
Among them, the zero-bias conductance peaks with point-contact
spectroscopy~\cite{Sasaki;Ando:2011,Kirzhner} can be strong evidence for
the topological superconductivity, although identifying the gap-function
type is an unsettled issue; the scanning tunneling spectroscopy
measurement~\cite{Levy;Stroscio:2013} shows that in 
$\mbox{Cu}_{0.2}\mbox{Bi}_{2}\mbox{Se}_{3}$ the density of states (DOS)
at the Fermi level is a fully-gapped behavior, leading to a conventional
(presumably, non-topological) superconducting state. 
Furthermore, the study about topological superconductors includes
different physical areas, such as color superconductivity in quark
matter.~\cite{Nishida:2010} 

Impurity effects are typically studied for
examining unconventional properties of superconductivity.~\cite{Kopnin:2001,Hirschfeld;Woelfe:1986,SchmittRink;Varma:1986,Hotta:1993,Preosti;Muzikar:1996} 
Assessing the impurity effects of topological superconductors is
desirable for not only finding the unconventional features, but also
developing a theory of nano-scale
devices composed of these materials. 
The authors~\cite{Nagai;Ota;Machida:2013} studied the impurity effects
in a two-dimensional topological superconductor with the Zeeman
term.~\cite{Sato;Fujimoto:2010}
Remarkably, the superconducting transition temperature
for a certain magnetic-field domain is significantly affected by
non-magnetic impurities, although the pair potential has an isotropic
$s$-wave character. 
The reduction of $T_{\rm c}$ is linked with the occurrence of surface
edge modes. 
This result arises a question. 
Do non-magnetic impurities diminish the superconductivity in other
models, especially systems with time-reversal symmetry? 

In this paper, we study the impurity effects in a three-dimensional
topological superconductor, with time-reversal symmetry and $s$-wave
on-site pairing. 
Using a mean-field Hamiltonian and a self-consistent $T$-matrix approach
for impurity scattering, we calculate the DOS and
examine the presence of in-gap states, with two kinds of fully-gapped
states, i.e., an even-parity state and an odd-parity one. 
We find that all the results are summarized well by a single material
variable. 
This variable characterizes \textit{relativistic} effects in the
Dirac Hamiltonian for normal part of this model. 
When the system is in non-relativistic regime, the odd-parity state is
suffered from non-magnetic impurities. 
In contrast, in relativistic regime, no low-energy excitation is induced 
by non-magnetic impurity; the topological superconductivity is robust
against non-magnetic impurities. 
To understand these behaviors, we derive an effective Hamiltonian in
two limiting cases. 
In the non-relativistic limit, the effective Hamiltonian
for the odd-parity gap function describes $p$-wave superconductivity. 
Since the Anderson's theorem is violated for $p$-wave symmetry, the
non-magnetic impurities induce low-energy in-gap states, and as a result
the superconducting state is fragile. 
In the ultra-relativistic limit, the effective Hamiltonian is the same as
the $s$-wave mean-field Hamiltonian, irrespective of the gap-function
types. 
In contrast to the non-relativistic limit, the Anderson's theorem
leads to the protection of the superconductivity from non-magnetic
impurities. 
To sum up, the full-gap states in this topological superconductor have 
%Consequently, we find that this three-dimensional topological
%superconductor has 
two aspects, $p$- and $s$-wave characters, depending
on the weight of relativistic effects in the normal-state Dirac
Hamiltonian. 

The paper is organized as follows. 
Section \ref{sec:formulation} shows the model and our approach for
impurity scattering in superconductivity. 
We also define an indicator of relativistic effects. 
Section \ref{sec:results} is main part of this paper. 
We show the DOS for two different gap functions, either even parity or
odd parity, changing the indicator of relativistic effects. 
The results are concisely summarized in Table \ref{table:1}. 
Section \ref{sec:summary} is devoted to summary.

\section{Formulation}
\label{sec:formulation}
\subsection{Model}
We study a model of a topological superconductor in a
three-dimensional system, with mean-field approximation. 
Typically, this model is used for examining Copper-doped
$\mbox{Bi}_{2}\mbox{Se}_{3}$ in literature~\cite{Fu;Berg:2009,Sasaki;Ando:2011,Hao;Lee:2011,Yamakage;Tanaka:2012,Nagai;Machida:2012,Mizushima;Tanaka:2013,Nagai;Nakamura;Machida:1211.0125,Nagai;Nakamura;Machida:1305.3025,Black-Schaffer}. 
The parent compound is considered to be a
three-dimensional strong topological insulator. 
The notable feature is the band structure around the $\Gamma$
point~\cite{Zhang:Zhang:2009}. 
The effective Hamiltonian with a strong spin-orbit coupling around
$\Gamma$ point  
is equivalent to that of the massive Dirac Hamiltonian with the negative
Wilson mass term. 
This negative term creates an insulator with a topological twist,
in terms of the Kane-Mele $Z_{2}$
invariant~\cite{Kane;Mele:2005}.
A low-energy effective theory of
$\mbox{Cu}_{x}\mbox{Bi}_{2}\mbox{Se}_{3}$ is derived, with the chemical
potential located in the conduction
band.~\cite{Fu;Berg:2009,Hao;Lee:2011,Sasaki;Ando:2011}

The mean-field Hamiltonian is 
\begin{equation}
H = \frac{1}{2} \sum_{\veck} \bm{\Psi}_{\veck}^{\dagger} 
\calH(\veck)
\bm{\Psi}_{\veck}. 
\label{eq:Hamiltonian_MFA}
\end{equation}
The $8$-component column vector $\bm{\Psi_{\veck}}$ has  the
electron annihilation operators $c_{\veck,\alpha,s}$ in the upper
$4$-component block (particle space) and the electron creation
operators $c^{\dagger}_{-\veck,\alpha,s}$ in the lower $4$-component
block (hole space), with momentum $\veck$, orbital $\alpha\,(=1,\,2)$,
and spin $s\,(=\uparrow,\,\downarrow)$. 
The arrangement of the upper $4$-component block is 
\(
\,^{\rm t}(c_{\veck,1,\uparrow},\,c_{\veck,1,\downarrow}, \,
c_{\veck,2,\uparrow},\,c_{\veck,2,\downarrow} )
\). 
The lower one is written in a similar manner. 
The Bogoliubov-de Gennes (BdG)
Hamiltonian~\cite{Hao;Lee:2011,Sasaki;Ando:2011} is an $8\times 8$
hermitian matrix,
\begin{equation}
\calH(\veck) 
=
\left(
\begin{array}{cc}
h_{0}(\veck) & \Delta_{\rm pair}(\veck)\\
\Delta_{\rm pair}^{\dagger}(\veck) & -h_{0}^{\ast}(-\veck)
\end{array}
\right),
\end{equation}
with the normal-state Hamiltonian 
\begin{equation}
h_{0}(\veck)
=
\varepsilon(\veck)
+ 
d_{0}(\veck)\gamma^{0}
+
\sum_{i=1}^{3}d_{i}(\veck)\gamma^{0}\gamma^{i},
\end{equation}
and the $4\times 4$ pairing potential matrix $\Delta_{\rm pair}$. 
The $4\times 4$ complex matrices $\gamma^{\mu}$ ($\mu=0,\,1,\,2,3$) are 
the Gamma matrices in the Dirac basis~\cite{Itzykson;Zuber:2005}. 
Using the orbital $2 \times 2$ Pauli matrices $\sigma^{i}$ and the spin
$2 \times 2$ Pauli 
matrices $s^{i}$, we have 
\mbox{$\gamma^{0} = \sigma^{3}\otimes \openone_{2}$} and 
\mbox{$\gamma^{i} = i\sigma^{2}\otimes s^{i}$}.

We summarize the properties of the normal part $h_{0}$ and
the superconducting part $\Delta_{\rm pair}$ in the BdG Hamiltonian. 
First, we focus on the normal part. 
We use the following formulae~\cite{Sasaki;Ando:2011}. 
For the diagonal block, we have 
\(
\varepsilon
=
-\mu
+
\bar{D}_{1} \epsilon_{c}(\veck)
+
(4/3)\bar{D}_{2} \epsilon_{\bot}(\veck)
\) 
and 
\(
d_{0}
=
M_{0}
- \bar{B}_{1}\epsilon_{c}(\veck)
- 
(4/3) \bar{B}_{2} \epsilon_{\bot}(\veck)
\), with 
\(
\epsilon_{c} = 2 -2 \cos(k_{z})
\), 
\(
\epsilon_{\bot} 
= 3 - 2\cos(\sqrt{3}k_{x}/2) \cos(k_{y}/2) - \cos(k_{y})
\). 
The off-diagonal block has contributions from the spin-orbit
couplings, 
\(
d_{1}
=
(2/3)\bar{A}_{2} 
\sqrt{3} 
\sin ( \sqrt{3}k_{x}/2 )
\cos ( k_{y} / 2 )
\), 
\(
d_{2}
=
(2/3)\bar{A}_{2} [
\cos ( \sqrt{3}k_{x} /2 )
\sin ( k_{y} /2 ) 
+
\sin ( k_{y} ) 
]
\), 
and 
\(
d_{3}
=
\bar{A}_{1} \sin(k_{z})
\). 
The material parameters ($\bar{D}_{1}$, $\bar{D}_{2}$, $M_{0}$,
$\bar{B}_{1}$, $\bar{B}_{2}$, $\bar{A}_{2}$, $\bar{A}_{1}$) and the
chemical potential $\mu$ will be set, in a manner compatible with
literature.~\cite{Sasaki;Ando:2011}

Next, we turn into the pairing potential $\Delta_{\rm pair}$. 
This $4\times 4$ matrix must fulfill the relation 
\mbox{
\(
\,^{\rm t}\Delta_{\rm pair}(-\veck) = -\Delta_{\rm
pair}(\veck)
\)},  
owing to the fermionic property of $c_{\veck,\alpha,s}$ and
$c_{\veck,\alpha,s}^{\dagger}$.~\cite{vanHemmen:1980} 
In this paper, we study the case when $\Delta_{\rm pair}$ has no
dependence on $\veck$. 
Using the above two properties, we have six possible
superconducting classes.~\cite{Hao;Lee:2011} 
They are classified by a Lorentz-transformation property.~\cite{Nagai;Machida:2012,Nagai;Nakamura;Machida:1211.0125,Nagai;Nakamura;Machida:1305.3025,Goswami} 
To see this point, we expand $\Delta_{\rm pair}i\gamma^{2}$ by a complete
orthogonal system (a set of sixteen complex matrices) built up by the
Gamma matrices and the identity.~\cite{Itzykson;Zuber:2005}  
A straightforward calculation leads to 
\begin{equation}
\Delta_{\rm pair} i\gamma^{2}
=
i\gamma^{0}
\bigg(
f^{\rm s}
+
f^{\rm ps} \gamma^{5}
+
\sum_{\mu=0}^{3}f^{\rm pv}_{\mu}\gamma^{\mu}
\bigg) \gamma^{5}, \label{eq:pair_expansion}
\end{equation}
with 
\(
\gamma^{5} 
= 
i\gamma^{0}\gamma^{1}\gamma^{2}\gamma^{3}
\). 
The possible non-zero coefficients are related to a scalar 
($f^{\rm s}$), a pseudo scalar ($f^{\rm ps}$), and a polar vector
($f^{\rm pv}_{\mu}$). 
Since the spatial inversion in this system~\cite{Sato:2010,note_P} is
defined by $\calP = \sigma^{3}\otimes \openone_{2}= \gamma^{0}$ and
transforms $\Delta_{\rm pair}$ into 
\(
\calP^{\dagger} \Delta_{\rm pair} \calP^{\ast}
\), the parity of the six superconducting
gaps~\cite{Fu;Berg:2009,Sato:2010} is described by 
\(
f^{\rm ps} \to - f^{\rm ps}
\), 
\(
f^{\rm pv}_{i} \to - f^{\rm pv}_{i}
\), 
\(
f^{\rm s} \to f^{\rm s}
\), and 
\(
f_{0}^{\rm pv} \to f^{\rm pv}_{0}
\). 
Therefore, the odd-parity superconductivity emerges when either 
$f^{\rm ps}$ or $f^{\rm pv}_{i}$ are not zero, and the others vanish.  
This superconducting order is characterized
topologically.~\cite{Fu;Berg:2009,Sato:2010} using an analogy with the 
$Z_{2}$ invariant for topological insulators with spatial inversion
symmetry.~\cite{Fu;Kane:2007}

\subsection{Indicator of relativistic effects}
We show a primary variable in this paper, to characterize the impurity
effects in the present model. 
Let us examine normal-state part $h_{0}$ closely. 
We find that the mass $M_{0}$ and the spin-orbit couplings
($\bar{A}_{2}$ and $\bar{A}_{1}$) are related to a low-energy behavior
of $h_{0}$. 
Taking $|\veck| \to 0$, we find that $h_{0}$ corresponds to the massive
Dirac equation with anisotropy along the $z$-axis, 
\begin{align}
h_{0} 
\approx - \mu + 
\gamma^{0}
[
M_{0}
+ 
\bar{A}_{2} (k_{x}\gamma^{1} + k_{y}\gamma^{2} ) 
+
\bar{A}_{1}k_{z}\gamma^{3}
]
+
\calO(|\veck|^{2})
. \label{eq:h0}
\end{align} 
Therefore, we may define an indicator of relativistic effects in $h_{0}$
as~\cite{note:A2}
\begin{equation}
 \beta = \frac{\bar{A}_{2} k_{F} }{|M_{0}|}, \label{eq:beta}
\end{equation}
with $\bar{A}_{2} k_{F} =  \sqrt{\mu^{2} - M_{0}^{2}}$.
The Dirac Hamiltonian has two distinct behaviors, depending on $\beta$;
a non-relativistic limit ($\beta \to 0$, i.e., a large-mass limit) and a
ultra-relativistic limit ($\beta \to \infty$, i.e., a massless limit).  
This variable plays a central role for understanding the impurity
effects.

\subsection{Formalism for impurity scattering: Self-consistent
  $T$-matrix approach}
\label{subsec:Tmatrix}
The effects of impurity scattering are taken as self-energy, with the
self-consistent $T$-matrix approximation. 
We consider the randomly distributed non-magnetic (magnetic)
impurity potentials [e.g., 
$V(\bm{r}) = \sum_{i} \delta(\bm{r}-\bm{r}_{i}) V^{\rm NM(M)}$]. 
The $T$-matrix for randomly distributed non-magnetic (magnetic)
impurities~\cite{Mahan:2000}  is 
\begin{equation}
T(\Omega) 
= \left[ 
1 - 
V^{\rm NM(M)} 
\frac{1}{N}\sum_{\veck} G_{\veck}(\Omega)
\right]^{-1} V^{\rm NM(M)} , 
\label{eq:tmatrix}
\end{equation}
with 
\begin{align}
V^{\rm NM} &= V_{0}
\left(\begin{array}{cc}
\openone_{2} \otimes \openone_{2}
 & 0 \\
0 & -\openone_{2} \otimes \openone_{2}
\end{array}\right),  \label{eq:nonmag} \\
V^{\rm M} &= V_{0}
\left(\begin{array}{cc}
\openone_{2} \otimes s^{3}
 & 0 \\
0 & -\openone_{2} \otimes s^{3}
\end{array}\right), \label{eq:mag}
\end{align}
where $N$ is the number of meshes in momentum space. 
The Green's function is  
\begin{align}
G_{\veck}(\Omega) 
=
\frac{1}{\Omega - \calH(\veck) - \Sigma(\Omega)}
=
\left(
\begin{array}{cc}
g_{\veck}(\Omega) & f_{\veck}(\Omega)  \\
f^{\dagger}_{\veck}(\Omega) & \bar{g}_{\veck}(\Omega)
\end{array}
\right), \label{eq:G}  
\end{align}
with the self-energy
\begin{equation}
\Sigma(\Omega) 
= n_{\rm imp} T(\Omega)-n_{\rm imp}V^{\rm NM(M)}. \label{eq:S}
\end{equation}
The impurity concentration is written by $n_{\rm imp}$. 
The second term in Eq.~(\ref{eq:S}) corresponds to the renormalization
of the chemical potential $\mu$, since in our calculations $\mu$ is
fixed. 
If one self-consistently calculates $\mu$ with fixed particle number,
this term is not needed. 
Solving Eqs.~(\ref{eq:tmatrix}), (\ref{eq:G}), and (\ref{eq:S})
self-consistently, we obtain the density of states (DOS) 
\begin{equation}
N(E) 
= -\frac{1}{2\pi} \frac{1}{N}\sum_{\veck}\,\tr[ {\rm Im}\, g_{\veck}(E)], 
\end{equation}
with given $\mu$ and gap amplitude $\Delta$. 
We will examine the impurity effects in topological superconductivity,
via the energy dependence of the DOS. 
We can find that the main results do not change, when
performing the full self-consistent calculations (i.e., the $T$-matrix
formula with the gap equation). 

\begin{table*}[t]
\caption{Summary of robustness against non-magnetic impurities, with
 different types of gap functions and $\beta$. 
The quantity $\beta$ characterizes the weight of relativistic effects [See
 Eq.~(\ref{eq:beta})]. 
When $\beta < 1$, the system is in non-relativistic regime, whereas
 when $\beta > 1$, the system is in relativistic regime. 
In the second column, according to
 Refs.~\onlinecite{Fu;Berg:2009,Sato:2010}, even parity indicates a
 non-topological state, while odd parity means a topological state. 
The fifth column shows the density of states without impurities at
 $E=3\Delta$, to estimate normal-state contributions. 
The last column shows the corresponding figures.}
\label{table:1}
\begin{ruledtabular}
\begin{tabular}{cccccccc}
Gap type &
Parity &
Relativity ($\beta$) & 
Robustness & 
Density of states (at $E=3\Delta$) & 
Figure \\
scalar & even &
0.88  &
robust &
$ 0.04 $ & Figure \ref{fig:type1}  \\
pseudo-scalar & odd & 
0.88 &
fragile &
0.04 &  Figure \ref{fig:type206}   \\
pseudo-scalar & odd & 
2.83 &
robust &
0.008 &  Figure \ref{fig:type201}   \\
pseudo-scalar & odd & 
2.83 &
robust &
0.04 & Figure \ref{fig:type2013v}  
\end{tabular}
\end{ruledtabular} 
\end{table*}

\section{Results}
\label{sec:results}
We study the impurity effects of both even- and odd-parity gaps,
checking in-gap states at low-energy (less than gap amplitude) region in
the DOS. 
All the results are summarized in Table \ref{table:1}. 
For even parity, we focus on the scalar gap, 
$f^{\rm s}$. 
Using the notations in Sasaki \textit{et al.},~\cite{Sasaki;Ando:2011}
we find that 
\(
\Delta^{11}_{\uparrow\downarrow}
=
- \Delta^{11}_{\downarrow\uparrow}
\), 
\(
\Delta^{22}_{\uparrow\downarrow}
=
- \Delta^{22}_{\downarrow\uparrow}
\), 
\(
\Delta^{11}_{\uparrow\downarrow}
=
\Delta^{22}_{\uparrow\downarrow}
\), and the other components of the pairing potential matrix are zero. 
The gap amplitude is 
$\Delta = |f^{\rm s}|=|\Delta^{11}_{\uparrow\downarrow}|$. 
In terms of Hao and Lee,~\cite{Hao;Lee:2011} this state correspond to
even-parity intra-orbital singlet pairing. 
For odd parity, we study the pseudo-scalar gap, $f^{\rm ps}$. 
In a similar manner to $f^{\rm s}$, we find that 
\(
\Delta^{12}_{\uparrow\downarrow}
=
- \Delta^{12}_{\downarrow\uparrow}
\), 
\(
\Delta^{21}_{\uparrow\downarrow}
=
- \Delta^{21}_{\downarrow\uparrow}
\), 
\(
\Delta^{12}_{\uparrow\downarrow}
=
\Delta^{21}_{\uparrow\downarrow}
\), and the others are zero.
The gap amplitude in this case is 
$\Delta = |f^{\rm ps}| = |\Delta^{12}_{\uparrow\downarrow}|$.  
%This corresponds to odd-parity inter-orbital
%singlet in literature.~\cite{Hao;Lee:2011}  
Both $f^{\rm s}$ and $f^{\rm ps}$ have a full-gap feature in their
spectral functions.~\cite{Sasaki;Ando:2011} 

Let us summarize the parameter sets for numerically calculating the
DOS in this paper.  
The $\veck$-mesh size is $256 \times 256 \times 256$. 
We focus on a unitary-like scattering model with 
$V_{0} = 10\,\mbox{eV}$ and $n_{\rm imp} = 0.005$, to study a case
that the superconducting pair is broken drastically. 
The gap amplitude is $\Delta = 0.1\,\mbox{eV}$ for both even- and
odd-parity superconducting states. 
The material parameters for most of the calculations are set
by 
\(
(\bar{D}_{1},\, \bar{D}_{2},\, \bar{B}_{1},\, \bar{B}_{2},\,
\bar{A}_{1},\, \bar{A}_{2})
= (0,\, 0,\, -0.5,\, -0.75,\, 1,\, 1.5)
\), in unit of $\mbox{eV}$. 
When we choose different values, we will show them explicitly. 
The unit of energy is $\mbox{eV}$ throughout this paper, unless
otherwise noted. 
These parameters are the same as the ones in
Ref.~\onlinecite{Sasaki;Ando:2011}. 
In the subsequent parts, the relativity indicator $\beta$ is tuned,
changing $M_{0}$ with 
\(
|\mu| - |M_{0}| = 0.2\, \mbox{eV}
\). 
This difference is the same value as in
Ref.~\onlinecite{Sasaki;Ando:2011}. 

\subsection{Scalar gap (non-topological superconductivity)}
\begin{figure}[thb]
\begin{center}
\begin{tabular}{p{ \columnwidth}} %p{0.5 \columnwidth}}%  p{28mm}}
\resizebox{ \columnwidth}{!}{\includegraphics{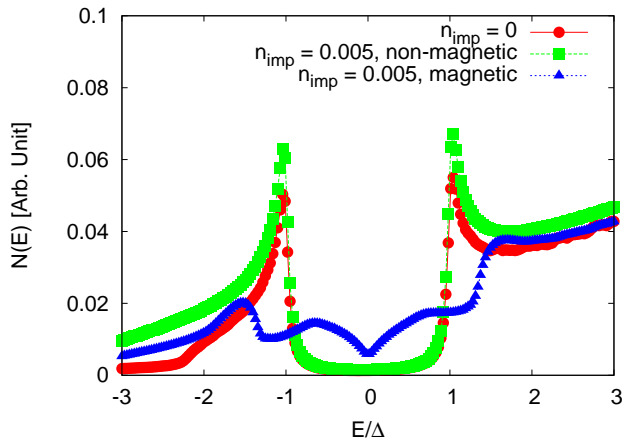}} 
\end{tabular}
\end{center}
\caption{\label{fig:type1} 
(Color online) Energy dependence of the density of states $N(E)$ in the
 scalar gap (a non-topological superconductor), with different types of
 impurities. 
The red circle indicates the DOS without impurity scattering, whereas
 the green square and the blue triangle are the results for,
 respectively, non-magnetic and magnetic impurities. 
The horizontal axis is energy $E/\Delta$, with superconducting gap
 amplitude $\Delta = 0.1~\mbox{eV}$.} 
\end{figure}

We study the impurity effects of the scalar gap $f^{\rm s}$ (even
parity). 
We set $M_{0} = -0.6\,\mbox{eV}$ and $\mu = 0.8\,\mbox{eV}$ (i.e.,
$\beta \sim 0.88$).  
Figure \ref{fig:type1} shows that the non-magnetic impurities (green
square) do not induce in-gap states. 
This result corresponds to the consequence of the Anderson's
theorem~\cite{Kopnin:2001}. 
Figure \ref{fig:type1} also shows that the Anderson's theorem is broken
for the magnetic impurities (blue triangle). 
What is an important thing here is that the resultant properties are
completely equivalent to those of the conventional $s$-wave
superconductivity. 

%%%

\subsection{Pseudo-Scalar gap (odd-parity topological superconductivity)} 
\begin{figure}[htb]
\begin{center}
     \begin{tabular}{p{ \columnwidth}} %p{0.5 \columnwidth}}%  p{28mm}}
      \resizebox{ \columnwidth}{!}{\includegraphics{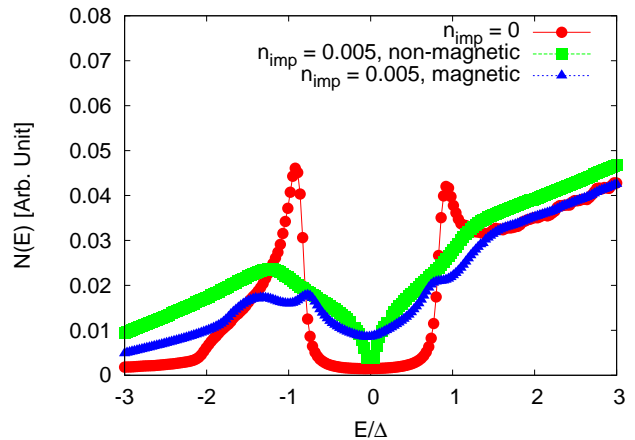}} %\\ %&
  %    \resizebox{\columnwidth}{!}{\includegraphics{Fig1b.eps}} 
    \end{tabular}
\end{center}
\caption{\label{fig:type206} 
(Color online) Energy dependence of the density of states $N(E)$ in a
 pseudo-scalar gap (an odd-parity topological superconductor) in a 
 non-relativistic region ($\beta \sim 0.88$), with different types of
 impurities. 
The legends are the same as in Fig.~\ref{fig:type1}. 
The horizontal axis is energy $E/\Delta$, with superconducting gap
 amplitude $\Delta = 0.1~\,\mbox{eV}$.} 
\end{figure}

Let us focus on the impurity effects of the pseudo-scalar gap (full-gap
  topological superconductivity). 
Firstly, we use the same mass and chemical potential as in
  Fig.~\ref{fig:type1}, $(M_{0},\mu) = (-0.6,0.8)$ 
(i.e., $\beta \sim 0.88$). 
Thus, we study the impurity effects in a non-relativistic region. 
Figure \ref{fig:type206} shows that both the non-magnetic (green square)
  and the magnetic (blue triangle) impurities induce in-gap states. 
The result for the non-magnetic impurities is similar to that in a
  two-dimensional $s$-wave topological
  superconductor.\,\cite{Nagai;Ota;Machida:2013}  
In the adopted parameters, the topological fully-gapped
  superconductivity is fragile against non-magnetic impurities. 

\begin{figure}[htb]
\begin{center}
     \begin{tabular}{p{ \columnwidth}} %p{0.5 \columnwidth}}%  p{28mm}}
      \resizebox{ \columnwidth}{!}{\includegraphics{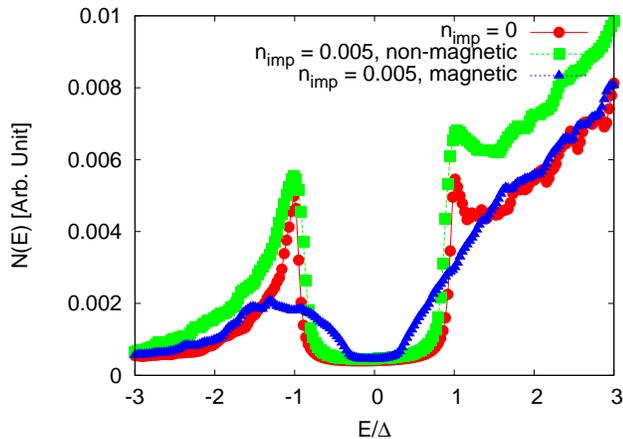}} %\\ %&
  %    \resizebox{\columnwidth}{!}{\includegraphics{Fig1b.eps}} 
    \end{tabular}
\end{center}
\caption{\label{fig:type201} 
(Color online) Energy dependence of the density of states $N(E)$ in a
 pseudo-scalar gap (an odd-parity topological superconductor) in a
 relativistic region ($\beta \sim 2.83$), with different types of
 impurities. 
The legends are the same as in Fig.~\ref{fig:type1}. 
The horizontal axis is energy $E/\Delta$, with superconducting gap
 amplitude $\Delta = 0.1~\,\mbox{eV}$. 
} 
\end{figure}

Secondly, we examine $\beta$-dependence of the robustness against
non-magnetic impurities. 
We set $(M_{0},\mu) = (-0.1,0.3)$. 
It means that $\beta \sim 2.83$ (a relativistic region).  
Figure \ref{fig:type201} shows that the non-magnetic impurities (green
square) does not induce any in-gap state. 
This remarkable effect is quite similar to the case for the
non-topological fully-gapped superconductivity (See
Fig.~\ref{fig:type1}). 
We note here that in a high energy domain (e.g., $E/\Delta \sim 3$) the DOS for the non-magnetic impurities is much different
from the one without impurities, compared to Figs.~\ref{fig:type1} and
\ref{fig:type206}.  
This originates from the fixed chemical potential and the resultant
filling change. 

\begin{figure}[tb]
\begin{center}
     \begin{tabular}{p{ \columnwidth}} %p{0.5 \columnwidth}}%  p{28mm}}
      \resizebox{ \columnwidth}{!}{\includegraphics{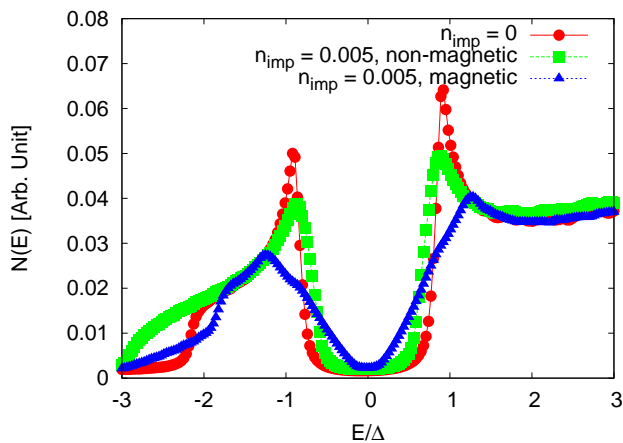}} %\\ %&
  %    \resizebox{\columnwidth}{!}{\includegraphics{Fig1b.eps}} 
    \end{tabular}
\end{center}
\caption{\label{fig:type2013v} 
(Color online) Energy dependence of the density of states $N(E)$ in a
 pseudo-scalar gap (an odd-parity topological superconductor) in a
 relativistic region ($\beta \sim 2.83$), with different types of
 impurities. 
The spin-orbit couplings are three times smaller than in
 Fig.~\ref{fig:type201}. 
The legends are the same as in Fig.~\ref{fig:type1}. 
The horizontal axis is energy $E/\Delta$, with superconducting gap
 amplitude $\Delta = 0.1~\mbox{eV}$. } 
\end{figure}

Now, we show that in the pseudo-scalar-type topological
superconductivity the robustness against non-magnetic impurities is
predominated by the relativity indicator $\beta$. 
Figures \ref{fig:type206} and \ref{fig:type201} show that 
the DOS at $E = 3\Delta$ for $(M_{0},\mu) = (-0.6,0.8)$
(Fig.~\ref{fig:type206}) is three times smaller than that for
$(M_{0},\mu) = (-0.1,0.3)$ (Fig.~\ref{fig:type201}). 
It indicates that the above calculations for a relativistic region
are performed when the normal-state DOS is relatively small. 
Intuitively, this small normal-state DOS can be an alternative origin to
reduce the effects of non-magnetic impurity scattering. 
To clarify the origin of the robustness, we examine a different model
parameter set, 
\(
(\bar{D}_{1},\,\bar{D}_{2},\,\bar{B}_{1},\,
\bar{B}_{2},\,\bar{A}_{1},\,\bar{A}_{2})
= (0,\,0,\,-0.5,\,-0.75,\,1/3,\,1.5/3)
\) 
and $(M_{0},\mu) = (-0.6,0.8)$. 
The relativity indicator is $\beta \sim 2.83$ (a relativistic region).  
We find that the spin-orbit couplings are three times smaller than in
the parameter set in Figs.~\ref{fig:type206} and \ref{fig:type201}. 
Figure \ref{fig:type2013v} shows that the DOS at $E = 3\Delta$ is
similar to Fig.~\ref{fig:type206}.
We stress that the robustness against non-magnetic impurities is the
same as in the previous relativistic region (See
Fig.~\ref{fig:type201}). 
Therefore, we claim that the relativity indicator $\beta$ characterizes
well the robustness against non-magnetic impurities in the full-gap
topological superconductivity. 

\subsection{Unified picture of impurity effects with a relativistic effect}
Now, we show deeper understanding of the present impurity effects, in terms
of relativity. 
To see this point, we derive an effective BdG Hamiltonian in a clean
case ($n_{\rm imp}=0$), with either the non-relativistic limit 
($\beta \to 0$) or the ultra-relativistic limit ($\beta \to \infty$). 
Since the relativity indicator $\beta$ appears in a low-energy behavior
in the normal-state Hamiltonian, we focus on linearized normal-state
Hamiltonian (\ref{eq:h0}). 
The $8 \times 8$ linearized BdG equations in the Dirac basis are 
\begin{align}
\left(\begin{array}{cccc}
M_{0}-\mu & \veck' \cdot {\bm s} & \Delta^{11} & \Delta^{12} \\
\veck' \cdot {\bm s} & -M_{0}-\mu & \Delta^{21} & \Delta^{22} \\
\Delta^{11 \dagger} & \Delta^{21 \dagger} & -M_{0}+\mu & \veck' \cdot {\bm s}^{\ast} \\
\Delta^{12 \dagger} & \Delta^{22 \dagger} & \veck' \cdot {\bm s}^{\ast} & M_{0}+\mu
\end{array}\right)
{\bm \phi} &= E {\bm \phi},
\end{align}
with 
\(
\,^{\rm t}\bm{\phi} 
= (^{\rm t}\bm{F}, \, ^{\rm t}\bm{G},\, ^{\rm t}\bm{F}^{\prime},\, 
^{\rm t}\bm{G}^{\prime}) 
\)
and 
\(
\veck^{\prime} 
= (\bar{A}_{2}k_{x},\, \bar{A}_{2}k_{y},\, \bar{A}_{1} k_{z})
\). 
We find that 
\(
\Delta^{11} = \Delta^{22} = i f^{\rm s} s^{2}
\)
and 
\(
\Delta^{12} = \Delta^{21} = i f^{\rm ps} s^{2}
\). 
%Even- and odd- parity $4 \times 4$ pair potentials shown in Eq.~(\ref{eq:pair_expansion}) in the Dirac basis can be respectively expressed as 
%\begin{align}
%\Delta^{\rm even} &\equiv 
%\left(\begin{array}{cc}
%\Delta^{11} & 0 \\
%0 & \Delta^{22}
%\end{array}\right), \\
%\Delta^{\rm odd} &\equiv 
%\left(\begin{array}{cc}
%0 & \Delta^{12} \\
%\Delta^{21} & 0
%\end{array}\right).
%\end{align}
%For example, the scalar gap function is expressed as 
%\begin{align}
%\Delta^{11} = \Delta^{22} = i f^{\rm s} s^{2},
%\end{align}
%and the pseudo scalar gap function is expressed as
%\begin{align}
%\Delta^{12} = \Delta^{21} = i f^{\rm ps} s^{2}.
%\end{align}

\subsubsection{Non-relativistic limit}
We consider the non-relativistic limit ($\beta \to 0$). 
Since the relativity indicator $\beta$ vanishes when $\mu=M_{0}$, we use
another variable  
\(
\mu^{\prime} = \mu - M_{0}
\), for examining the non-relativistic case. 
The non-relativistic behaviors of the Dirac equation are obtained by
expansion with respect to the order of $1/M_{0}$.  
In this approach~\cite{Itzykson;Zuber:2005}, the upper two components
of the Dirac spinor in the Dirac basis is regarded as large
components, whereas the others are small ones. 
Precisely speaking, the large components correspond to
the $+1$-eigenspace of $\gamma^{0}$ in the Dirac basis. 
Dropping parts relevant to the small components order by order,
projection onto the large components is achieved. 
In the BdG equations, we consider $\bm{F}$ and $\bm{F}^{\prime}$ to be
the large components. 
Then, we perform inverse-mass expansion, under the condition that 
$E/|M_{0}| \ll 1$, $\mu^{\prime}/|M_{0}| \ll 1$, and 
$\Delta/ |M_{0}| \ll 1$. 
The resultant equations are given as 
\begin{align}
\left(\begin{array}{cc}
\frac{\veck'^{2}}{2 M_{0}} - \mu' & \Delta_{\rm eff}(\veck') \\
\Delta^{\dagger}_{\rm eff}(\veck')& -\frac{\veck'^{2}}{2 M_{0}} + \mu' 
\end{array}\right)
\left(\begin{array}{c}
{\bm F} \\
{\bm F'}
\end{array}\right) &= E \left(\begin{array}{c}
{\bm F} \\
{\bm F'}
\end{array}\right), 
\end{align}
with 
\begin{align}
\Delta_{\rm eff}(\veck') 
%&= \frac{1}{2 M_{0}} \left[
% \veck' \cdot {\bm s} \Delta^{21} - \Delta^{12} \veck' \cdot {\bm s}^{\ast}
% \right]
&= \left[ f^{\rm s}  + 
\frac{f^{\rm ps}}{M_{0}}  (\veck^{\prime} \cdot \vecs)  \right] (i s^{2}). \label{eq:sp}
\end{align}
For even parity ($f^{\rm s}\neq 0$ and $f^{\rm ps}=0$), 
the above equations are equivalent to the BdG equations of the $s$-wave
superconductivity. 
This indicates that the non-topological fully-gapped superconductivity
is regarded as the conventional $s$-wave superconductivity, in the
non-relativistic limit. 
In contrast, when $f^{\rm s}=0$ and $f^{\rm ps}\neq 0$ (odd parity), 
the effective gap function (\ref{eq:sp}) is equivalent to a $p$-wave one. 
Therefore, the pseudo-scalar gap function in the non-relativistic limit
is fragile against non-magnetic impurities, since the Anderson's theorem
is not valid in the $p$-wave superconductivity. 
This $p$-wave like behavior is consistent with the results by a
quasi-classical approach~\cite{Nagai;Nakamura;Machida:1305.3025}. 
Two of the authors (YN and MM) showed that the pseudo-scalar gap is
equivalent to the spin-triplet order parameter, within the
quasi-classical approximation. 
In the spin-triplet order parameter, the $\vecd$-vector rotates
in momentum space (i.e., $\vecd(\veck) \propto \veck$). 
Therefore, the superconductivity is broken significantly via impurity
scattering. 
The induced DOS in the low energy region shown in Fig.~
\ref{fig:type206} can be explained by this correspondence.  
%In the topological fully-gapped superconductivity characterized as the pseudo-scalar gap $f^{\rm ps}$, 
%the $4 \times 4$ pair potential is expressed as 
%\begin{align}
%\Delta^{\rm ps} &= f^{\rm ps} 
%\left(\begin{array}{cc}
%0 & i s^{2} \\
%i s^{2} & 0
%\end{array}\right).
%\end{align}
%Thus, 
%the effective gap becomes 
%\begin{align}
%\Delta^{\rm ps}_{\rm eff}(\veck') &= \frac{f^{\rm ps}}{M_{0}}   \veck' \cdot {\bm s}  (i s^{2}). 
%\end{align}
%This effective gap function is equivalent to a $p$-wave gap function. 
%Thus, the pseudo-scalar gap function in the non-relativistic limit is fragile against non-magnetic impurities, since 
%the Anderson's theorem is not valid in $p$-wave superconductors.

\subsubsection{Ultra-relativistic limit}
We study the ultra-relativistic limit ($\beta \to \infty$). 
We use the Weyl basis, since the
normal-state Hamiltonian with $\beta \to \infty$ corresponds to the
massless Dirac equation (the Weyl equation).  
The chirarity (i.e., the eigenvalues of $\gamma^{5}$) is good
quantum number when $M_{0}\to 0$. 
Thus, in the Weyl basis $\gamma^{5}$ is a diagonal form (in the Dirac
basis $\gamma^{0}$ is diagonal). 
We will use over-line symbol to specify the quantities in the Weyl basis. 
The linearized BdG equations are 
\begin{align}
\left(\begin{array}{cc}
\bar{h}_{0} (\veck^{\prime}) & \bar{\Delta}_{\rm pair} \\
\bar{\Delta}_{\rm pair}^{\dagger} & - \bar{h}_{0}^{\ast}(-\veck')
\end{array}\right)
\bm{\bar{\phi}} &= E \bm{\bar{\phi}},
\end{align}
with 
\begin{align}
\bar{h}_{0} &=
\left(\begin{array}{cc}
-\veck^{\prime} \cdot {\bm s}-\mu  & M_{0} \\
M_{0} & \veck^{\prime} \cdot {\bm s} -\mu
\end{array}
\right),
\end{align}
and 
\(
^{t}\bm{\bar{\phi}} 
=
(^{t}\bm{\bar{F}},\,^{t}\bm{\bar{G}},\,
^{t}\bm{\bar{F}}^{\prime},\,^{t}\bm{\bar{G}}^{\prime})
\). 
The $4\times 4$ pairing potential matrix including the scalar and the
pseudo-scalar gaps is 
\begin{align}
\bar{\Delta}_{\rm pair} 
&= f^{\rm s}
\left(
\begin{array}{cc}
i s^{2}& 0 \\
0 &  i s^{2}
\end{array}
\right)
+
f^{\rm ps} 
\left(
\begin{array}{cc}
 - i s^{2}&0 \\
0 & i s^{2}
\end{array}
\right).
\end{align}
Taking $M_{0} \rightarrow 0 $, we find that the BdG equations
are divided into a left-handed sector ($\bm{\bar{G}}$ and
$\bm{\bar{G}}^{\prime}$) and a right-handed sector ($\bm{\bar{F}}$ and
$\bm{\bar{F}}^{\prime}$), for both the scalar and the pseudo-scalar
gaps. 
As for $\bm{\bar{G}}$ and $\bm{\bar{G}}^{\prime}$, we have
\begin{align}
\left(
\begin{array}{cc}
\veck^{\prime} \cdot \bm{s} -\mu & i s^{2} f^{\rm s(ps)} \\
-i s^{2} f^{\rm s(ps)\, \ast} & \veck' \cdot \bm{s}^{\ast} + \mu
\end{array}\right)
\left(\begin{array}{c}
\bm{\bar{G}} \\
\bm{\bar{G}}^{\prime}
\end{array}\right) &=
E \left(\begin{array}{c}
\bm{\bar{G}} \\
\bm{\bar{G}}^{\prime}
\end{array}
\right). 
%\\
%\left(\begin{array}{cc}
%-\veck' \cdot {\bm s} -\mu & \pm i s^{2} f^{\rm s(ps)} \\
%\mp i s^{2} f^{\rm s(ps) \ast} & -\veck' \cdot {\bm s}^{\ast} + \mu
%\end{array}\right)
%\left(\begin{array}{c}
%{\bm \bar{F}} \\
%{\bm \bar{F}'}
%\end{array}\right) &=
%E \left(\begin{array}{c}
%{\bm \bar{F}} \\
%{\bm \bar{F}'}
%\end{array}\right),
\end{align}
We can find the same expression for $\bm{\bar{F}}$ and
$\bm{\bar{F}}^{\prime}$, except for the sign factor in each block. 
Remarkably, both the scalar and the pseudo-scalar effective gap
functions are equivalent to the $s$-wave gap function.
Furthermore, we find that non-magnetic potential (\ref{eq:nonmag}) and
magnetic impurity potentials (\ref{eq:mag}) are decoupled between the
left-handed and the right-handed sectors in the Weyl basis. 
These results indicate that the present model in the ultra-relativistic
limit has a robust feature against non-magnetic impurities associated
with the Anderson's theorem, irrelevant with the parity.

\section{Summary}
\label{sec:summary}
In conclusion, we calculated the density of states and examined the
presence of in-gap states in the three-dimensional topological
superconductors described by the massive Dirac Hamiltonian, using a
self-consistent $T$-matrix approach.  
We found that the results are summarized well by the material variable
$\beta$, which characterize relativistic effects in $h_{0}$.   
In non-relativistic region, we find that the Anderson's theorem (i.e.,
the robustness against non-magnetic impurities ) is violated, since a
$p$-wave character involved in the topological superconducting order is
predominant. 
In ultra-relativistic region, the effective Hamiltonian is the same as
the $s$-wave Hamiltonian, irrespective of the gap-function types.  
Therefore, we claim that a three-dimensional topological
superconductor with time-reversal symmetry and $s$-wave on-site pairing
has either $p$- or $s$-wave characters, depending on the
relativistic effects in the normal-state Hamiltonian.

\section*{Acknowledgements}
We thank H. Nakamura for helpful discussions and comments. 
The calculations were performed using the supercomputing system PRIMERGY
BX900 at the Japan Atomic Energy Agency.  
This study has been supported by Grants-in-Aid for Scientific Research
from the Ministry of Education, Culture, Sports, Science and Technology
of Japan.


\begin{thebibliography}{99}
\bibitem{note:A2}
We consider that this indicator only depends on $\bar{A}_{2}$, not $\bar{A}_{1}$,   
since the in-plane spin-orbit coupling $\bar{A}_{2}$ is always larger than the inter-plane spin-orbit coupling $A_{1}$.
\end{thebibliography}
\end{document}